\documentclass[12pt]{article}
\usepackage{graphicx}
\usepackage{xcolor}

\newcommand\pubnumber{CMS-CR-2023/198}
\newcommand\pubdate{\today}

\def\institute{Institute of Experimental Particle Physics (ETP),\\
Karlsruhe Institute of Technology (KIT), 76131 Karlsruhe, GERMANY}
\def\authemail{\footnote{Contact: nils.faltermann@cern.ch}}

\def\Title#1{\begin{center} {\Large #1 } \end{center}}
\def\Author#1{\begin{center}{ \sc #1} \end{center}}
\def\Address#1{\begin{center}{ \it #1} \end{center}}

\newcommand\pubblock{\rightline{\begin{tabular}{l} \pubnumber\\
         \pubdate  \end{tabular}}}
\newenvironment{Abstract}{\begin{quotation}  }{\end{quotation}}
\newenvironment{Presented}{\begin{quotation} \begin{center}
             PRESENTED AT\end{center}\bigskip
      \begin{center}\begin{large}}{\end{large}\end{center} \end{quotation}}

\def\beq{\begin{equation}}
\def\eeq#1{\label{#1}\end{equation}}
\def\eeqn{\end{equation}}

\def\beqa{\begin{eqnarray}}
\def\eeqa#1{\label{#1}\end{eqnarray}}
\def\eeqan{\end{eqnarray}}

\let\bar=\overbar

\def\Dslash{\not{\hbox{\kern-4pt $D$}}}
\def\dslash{\not{\hbox{\kern-2pt $\del$}}}

\def\msb{{\bar{\ssstyle M \kern -1pt S}}}

\begin{document}
\begin{titlepage}
\pubblock

\vfill
\Title{Associated single top production (t+X)}
\vfill
\Author{Nils Faltermann\authemail~on behalf of the ATLAS and CMS Collaborations}
\Address{\institute}
\vfill
\begin{Abstract}
Single top quark production in association with vector bosons provides a unique way to probe the electroweak sector of the standard model at the Large Hadron Collider. In this talk the latest experimental results of the ATLAS and CMS Collaborations for these processes are presented.
\end{Abstract}
\vfill
\begin{Presented}
$16^\mathrm{th}$ International Workshop on Top Quark Physics\\
(Top2023), 24--29 September, 2023
\end{Presented}
\vfill
\let\thefootnote\relax\footnotetext{Copyright 2023 CERN for the benefit of the ATLAS and CMS Collaborations. Reproduction of this article or parts of it is allowed as specified in the CC-BY-4.0 license}
\end{titlepage}
\def\thefootnote{\fnsymbol{footnote}}
\setcounter{footnote}{0}

\section{Introduction}
With almost $140\,\mathrm{fb}^{-1}$ of proton-proton collision data collected at the Large Hadron Collider (LHC) between 2015 and 2018, both the ATLAS and CMS Collaborations~\cite{ATLAS:2008xda,CMS:2008xjf} are not only able to measure precisely single top quark production in the $t$ and $s$ channel, as well as through W-associated production, but also to search for single top quark production in association with additional neutral vector bosons. These rare processes allow to probe different fermion and boson couplings and, in case of deviations from the standard model (SM) predictions, could hint to physics beyond the SM.

\section{tZq production}
The associated production of single top quarks in the $t$ channel with a Z boson (tZq) allows to probe both the coupling of the Z boson to the top quark, as well as to the W boson. To be sensitive to this production mode the experiments are targeting only the Z boson decay into charged leptons and require $m_{\ell\ell} > 30 \,\mathrm{GeV}$ for the signal definition.

\begin{figure}[!h!tbp]
\centering
\includegraphics[width=0.31\textwidth]{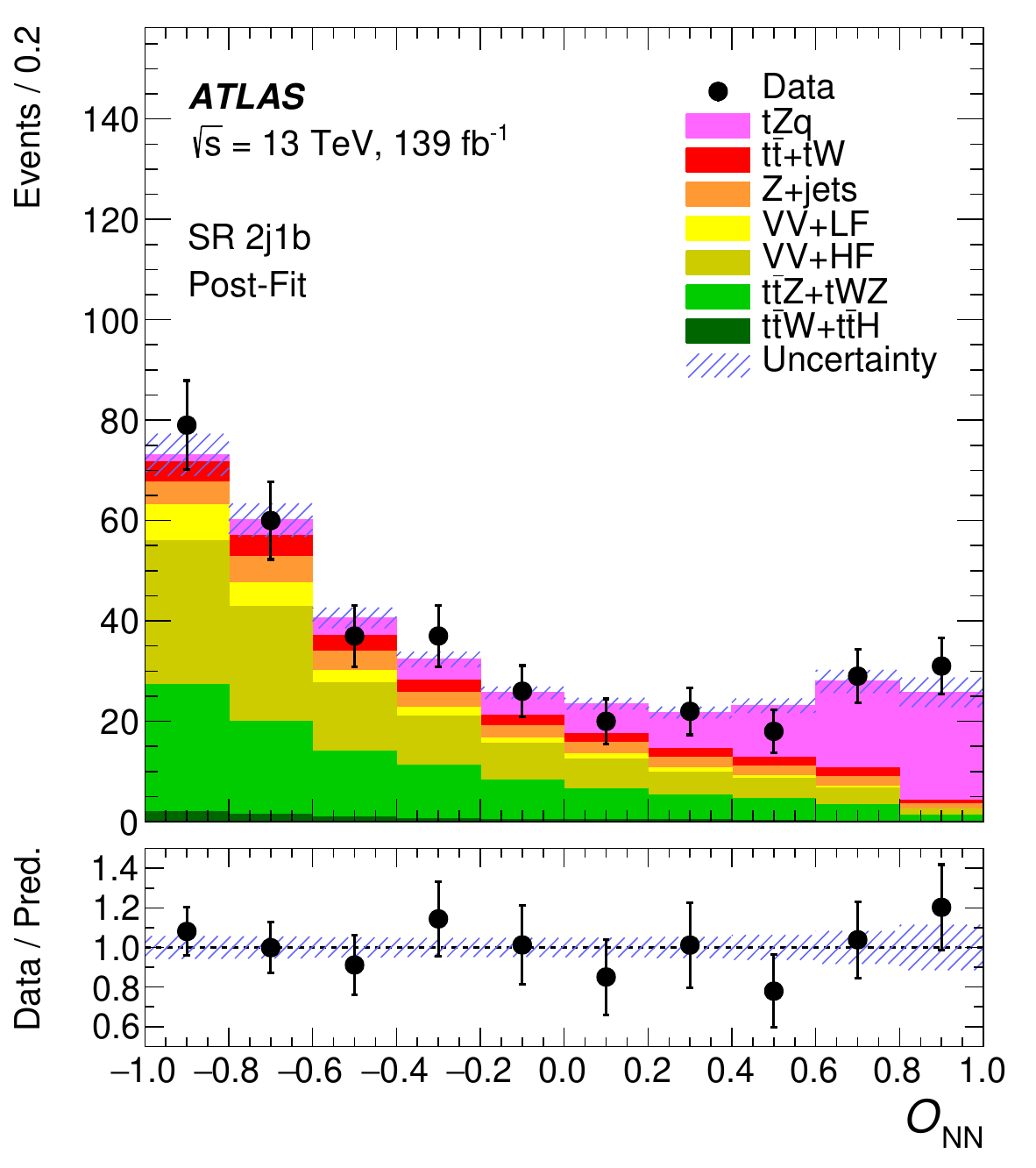}
\includegraphics[width=0.33\textwidth]{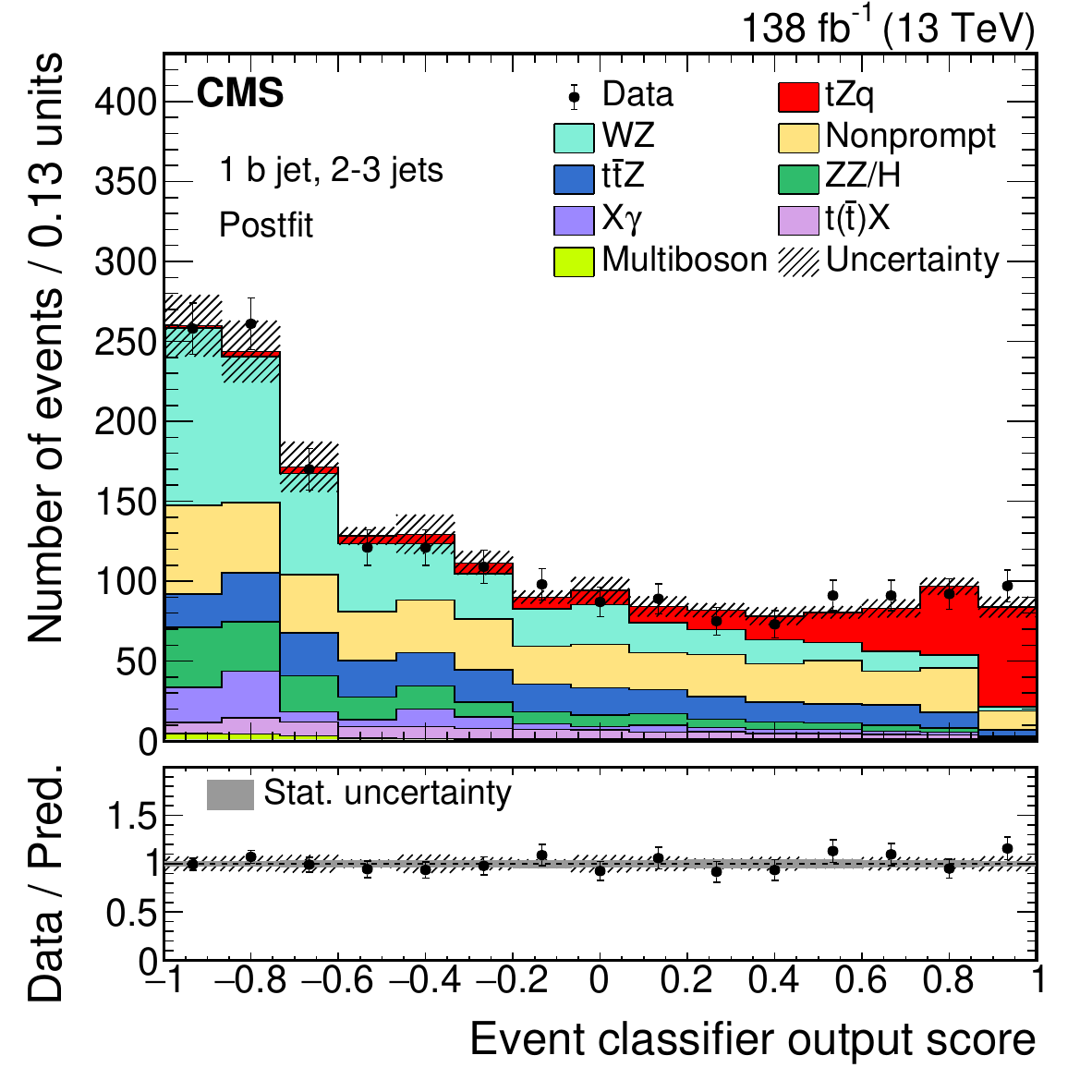}
\includegraphics[width=0.34\textwidth]{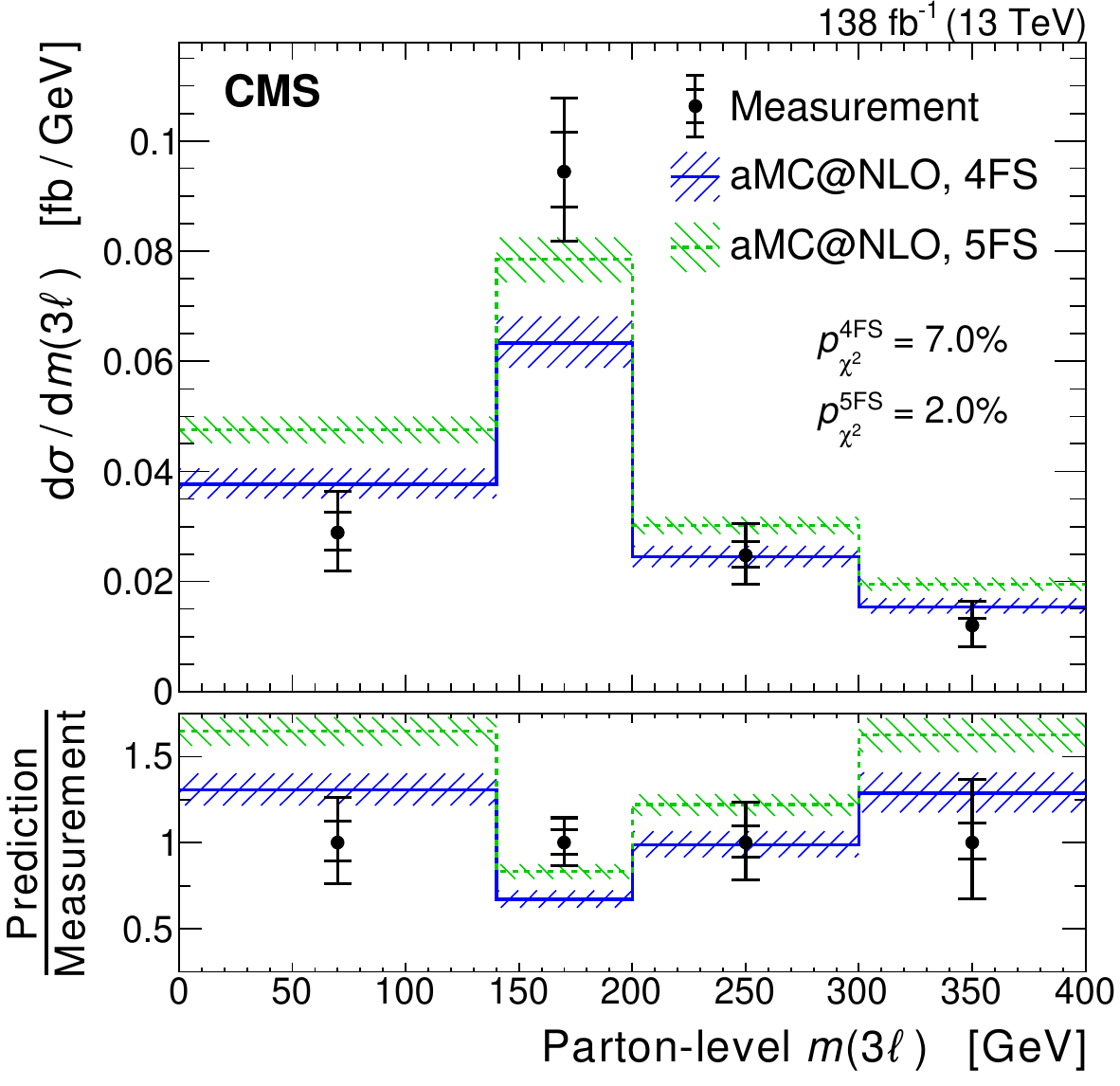}
\caption{Postfit distributions of the NN output score (left) and BDT output score (middle) in the measurement of tZq production by the ATLAS and CMS Collaborations, respectively~\cite{ATLAS:2020bhu,CMS:2021ugv}. In the latter analysis, the cross section is also measured differentially, shown here for the invariant mass of the three-lepton system at parton level (right).}
\label{fig:tzq}
\end{figure}

The ATLAS Collaboration has searched for tZq production using $139 \,\mathrm{fb}^{-1}$ of data~\cite{ATLAS:2020bhu}. Selected events require exactly three charged leptons (electrons or muons). Two of them need to be of opposite charge and the same flavor to target the Z boson decay, while the third lepton is assigned to the leptonic decay of a W boson from the subsequent decay of the top quark. Further event categorization is based on the number of jets and b-tagged jets in the event, assigning events either to one of two signal regions (SRs) or to different control regions (CRs). Background processes in this analysis constitute either of prompt leptons from diboson and top quark pair associated production with vector bosons (ttV) or the Higgs boson (ttH), or from nonprompt leptons. The background prediction is estimated from simulation and validated for the nonprompt leptons with an orthogonal selection using leptons from b hadron decays. To enhance the sensitivity to the signal process a neural network (NN) is trained. The signal is then extracted in a simultaneous maximum likelihood fit in all SRs and CRs and a significance of $>$5$\sigma$ is found with respect to the background-only hypothesis with a measured cross section for tZq production of
\beq
\sigma_{\mathrm{tZq}} = 97 \pm 13 \,\mathrm{(stat)} \pm 7 \,\mathrm{(syst)} \,\mathrm{fb} \,,
\eeq{eq:tzq_atlas}
which agrees with the SM prediction of $ \sigma_{\mathrm{tZq}}^{\mathrm{SM}} = 102 \,^{+5} _{-2} \,\mathrm{(scale+PDF)} \,\mathrm{fb} $ at next-to-leading order (NLO) in quantum chromodynamics (QCD). The most dominant systematic uncertainty in this measurement is the prompt lepton background estimation. The NN output distribution after the fit in one of the SRs is shown in Fig.~\ref{fig:tzq} (left).

The measurement of the CMS Collaboration with $138\,\mathrm{fb}^{-1}$ of data~\cite{CMS:2021ugv} follows a similar strategy for the event selection. The prompt lepton background is estimated from simulation, while the nonprompt lepton contribution is modeled directly from data in a nonprompt-enriched sideband. A boosted decision tree (BDT) is trained to maximize the signal sensitivity. The inclusive cross section is measured as
\beq
\sigma_{\mathrm{tZq}} = 87.9 \,^{+7.5} _{-7.3} \,\mathrm{(stat)} \,^{+7.3} _{-6.0} \,\mathrm{(syst)} \,\mathrm{fb} \,,
\eeq{eq:tzq_cms}
also in agreement with the prediction of $ \sigma_{\mathrm{tZq}}^{\mathrm{SM}} = 94.2 \,^{+1.9} _{-1.8} \,\mathrm{(scale)} \pm 2.5 \,\mathrm{(PDF)} \,\mathrm{fb} $ at NLO accuracy in QCD. The modeling of the signal process and the nonprompt background estimation are the dominating systematic uncertainties in this measurement. The postfit distribution of the BDT output score for one the SRs is shown in Fig.~\ref{fig:tzq} (middle). In addition to the inclusive cross section, the tZq production is also measured differentially for several observables. A likelihood-based method is used to unfold the kinematic distributions to parton and particle level, where they are compared to different theoretical predictions. Exemplary shown in Fig.~\ref{fig:tzq} (right) is the invariant mass of the three-lepton system unfolded to parton level.

\section{tq$\gamma$ production}
Analogously to the measurements of tZq production, the searches for single top quark production in association with a photon (tq$\gamma$) target the $t$-channel production mode. With the additional photon it is possible to probe the electric charge, as well as the electric and magnetic dipole moments of the top quark.

\begin{figure}[!h!tbp]
\centering
\includegraphics[width=0.45\textwidth]{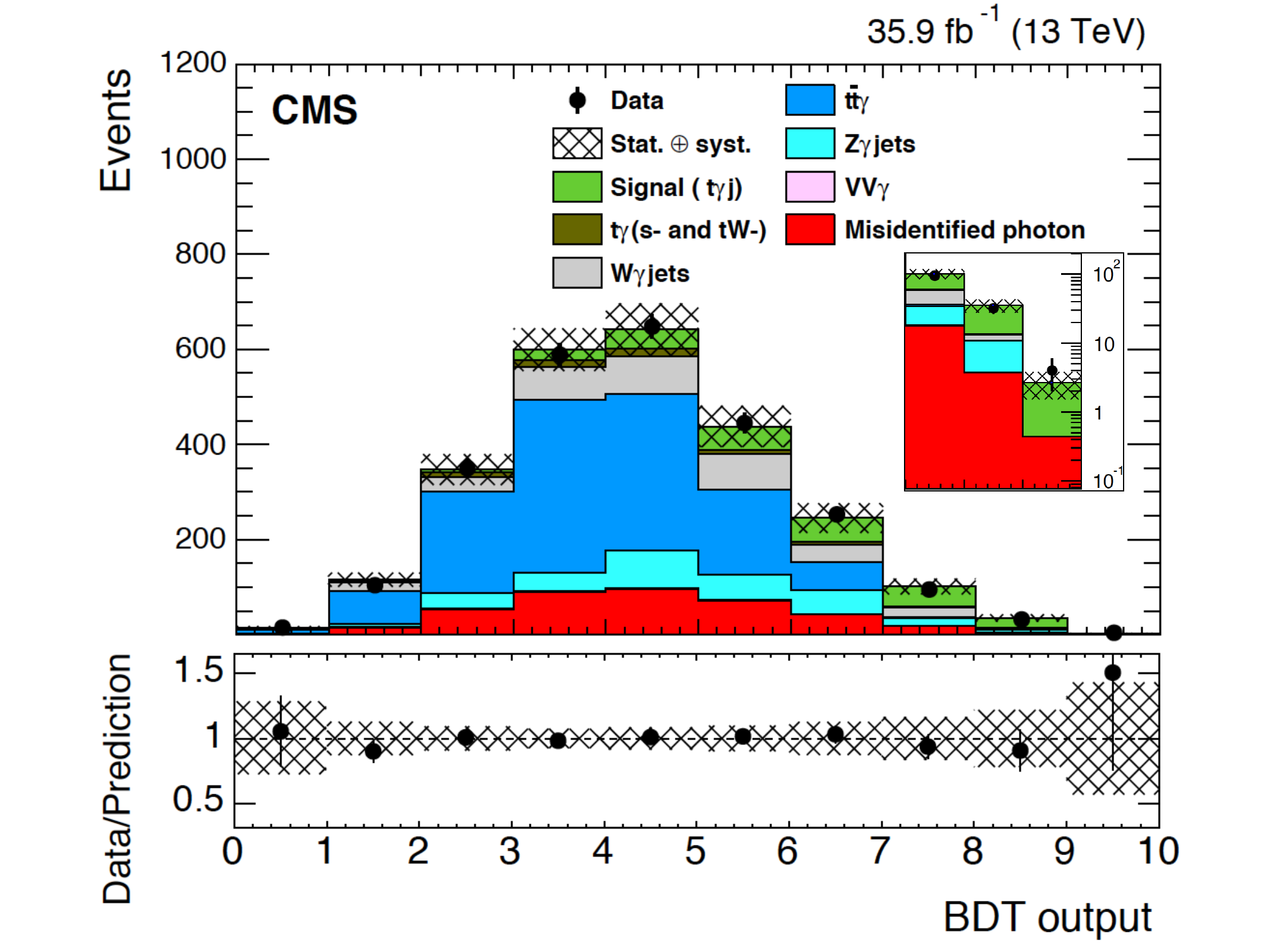}
\includegraphics[width=0.31\textwidth]{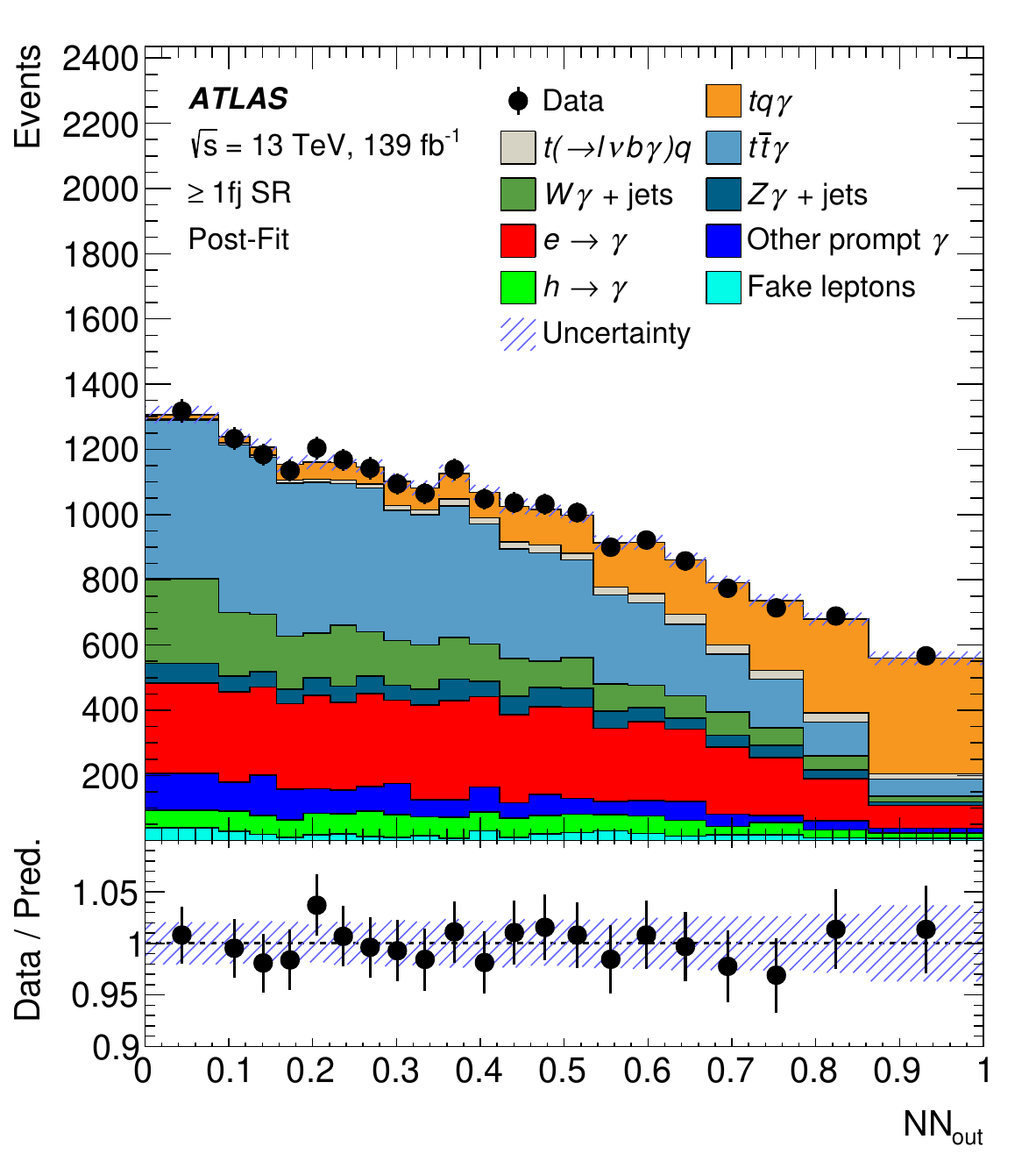}
\caption{Postfit distributions of the BDT output score (left) and NN output score (right) in the search for tq$\gamma$ production by the CMS and ATLAS Collaborations, respectively~\cite{CMS:2018hfk,ATLAS:2023qdu}.}
\label{fig:tqgamma}
\end{figure}

The CMS Collaboration has conducted a search for tq$\gamma$ production with $35.9 \, \mathrm{fb}^{-1}$ of data~\cite{CMS:2018hfk}. Events for the SR are selected based on the muonic decay of a W boson from the top quark decay and the presence of an isolated photon in the detector. A dedicated CR enriched in events with top quark pair plus photon production (tt$\gamma$) is defined to directly model this background process from data, while other background processes with prompt photons are estimated from simulation. Misidentified photons, either originating from true electrons or hadrons, are modeled from a sideband region and the normalization is estimated with an ABCD method. The output distribution of the BDT is fitted simultaneously in the SR and CR to extract a possible signal, while the tt$\gamma$ contribution is left freely floating. A signal is observed with a significance of $4.4\sigma$ ($3.0\sigma$ expected), thus being the first evidence for tq$\gamma$ production at the LHC. The measured fiducial cross section
\beq
\sigma_{\mathrm{tq}\gamma} \cdot \mathcal{B}(\mathrm{t\rightarrow\mu\nu b}) = 115 \pm 17 \,\mathrm{(stat)} \pm 30 \,\mathrm{(syst)} \,\mathrm{fb}
\eeq{eq:tqgamma_cms}
agrees with the prediction of $ \sigma_{\mathrm{tq}\gamma}^{\mathrm{SM}} \cdot \mathcal{B}(\mathrm{t\rightarrow\mu\nu b}) = 81 \pm 4 \,\mathrm{(scale+PDF)} \,\mathrm{fb} $ (NLO QCD) within the uncertainty of the measurement, which is systematically limited by the uncertainty in the jet energy scale and the signal modeling. The postfit distribution of the BDT output score is shown in Fig.~\ref{fig:tqgamma} on the left.

A search for tq$\gamma$ production is also carried out by the ATLAS Collaboration with $139\,\mathrm{fb}^{-1}$ of data~\cite{ATLAS:2023qdu}, targeting both electron and muon final states. Events are divided into two different SRs, depending on the presence of jet in forward direction of the detector. The prompt photon background contribution is estimated from simulation in dedicated CRs, while the contribution from misidentified photons is estimated with data-driven methods. Multiple NNs are trained to be sensitive to the signal, which is extracted from a combined maximum likelihood fit in all SRs and CRs. A tq$\gamma$ signal is observed with a significance of $9.1\sigma$ ($6.7\sigma$ expected), resulting in the first observation of this process at the LHC. Comparing the measured fiducial cross section of
\beq
\sigma_{\mathrm{tq}\gamma} \cdot \mathcal{B}(\mathrm{t\rightarrow\ell\nu b}) = 688 \pm 23 \,\mathrm{(stat)} \,^{+75} _{-71} \,\mathrm{(syst)} \,\mathrm{fb}
\eeq{eq:tqgamma_atlas_parton}
with the prediction at parton level of $ \sigma_{\mathrm{tq}\gamma}^{\mathrm{SM}} \cdot \mathcal{B}(\mathrm{t\rightarrow\ell\nu b}) = 515 \,^{+36} _{-42} \,\mathrm{(scale+PDF)} \,\mathrm{fb} $ at NLO in QCD results in a compatibility with the SM of $2.1\sigma$. At particle level, where contributions from photons coupling to the decay products of the top quark are also included, the fiducial cross section
\beq
\sigma_{\mathrm{tq}\gamma} \cdot \mathcal{B}(\mathrm{t\rightarrow\ell\nu b}) + \sigma_{\mathrm{t}(\rightarrow \ell\nu \mathrm{b} \gamma)\mathrm{q}} = 303 \pm 9 \,\mathrm{(stat)} \,^{+33} _{-32} \,\mathrm{(syst)} \,\mathrm{fb}
\eeq{eq:tqgamma_atlas_particle}
shows a similar tension with the predicted value of $ \sigma_{\mathrm{tq}\gamma}^{\mathrm{SM}} \cdot \mathcal{B}(\mathrm{t\rightarrow\ell\nu b}) + \sigma_{\mathrm{t}(\rightarrow \ell\nu \mathrm{b} \gamma)\mathrm{q}} = 217 \,^{+27} _{-15} \,\mathrm{(scale+PDF)} \,\mathrm{fb} $. Most dominant systematic uncertainties are the tt$\gamma$ modeling and the statistical power of the background simulation. The fit result for the NN distribution in one of the SRs is shown in Fig.~\ref{fig:tqgamma} (right).

\section{tWZ production}
\begin{figure}[!h!tbp]
\centering
\includegraphics[width=0.325\textwidth]{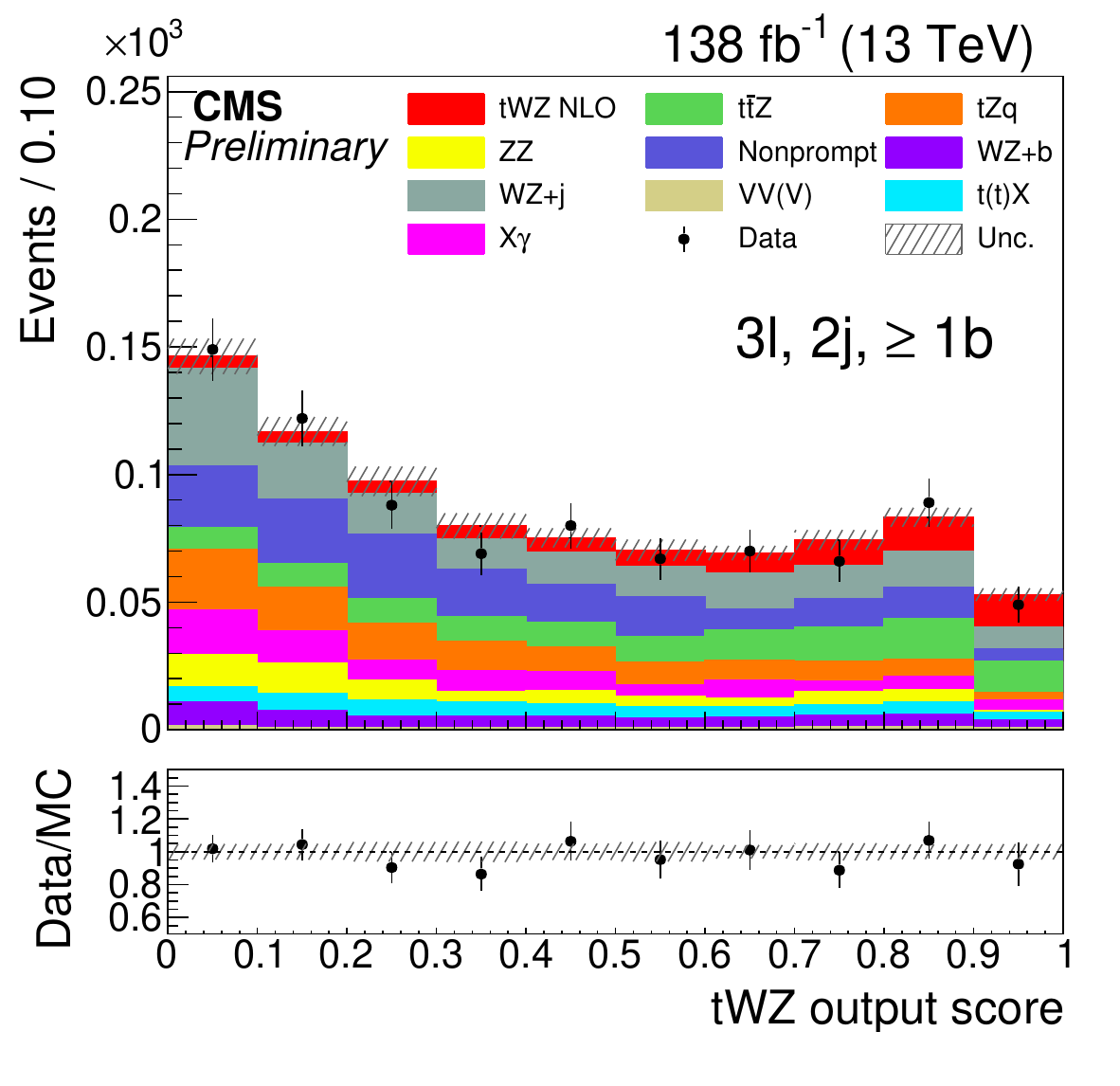}
\includegraphics[width=0.325\textwidth]{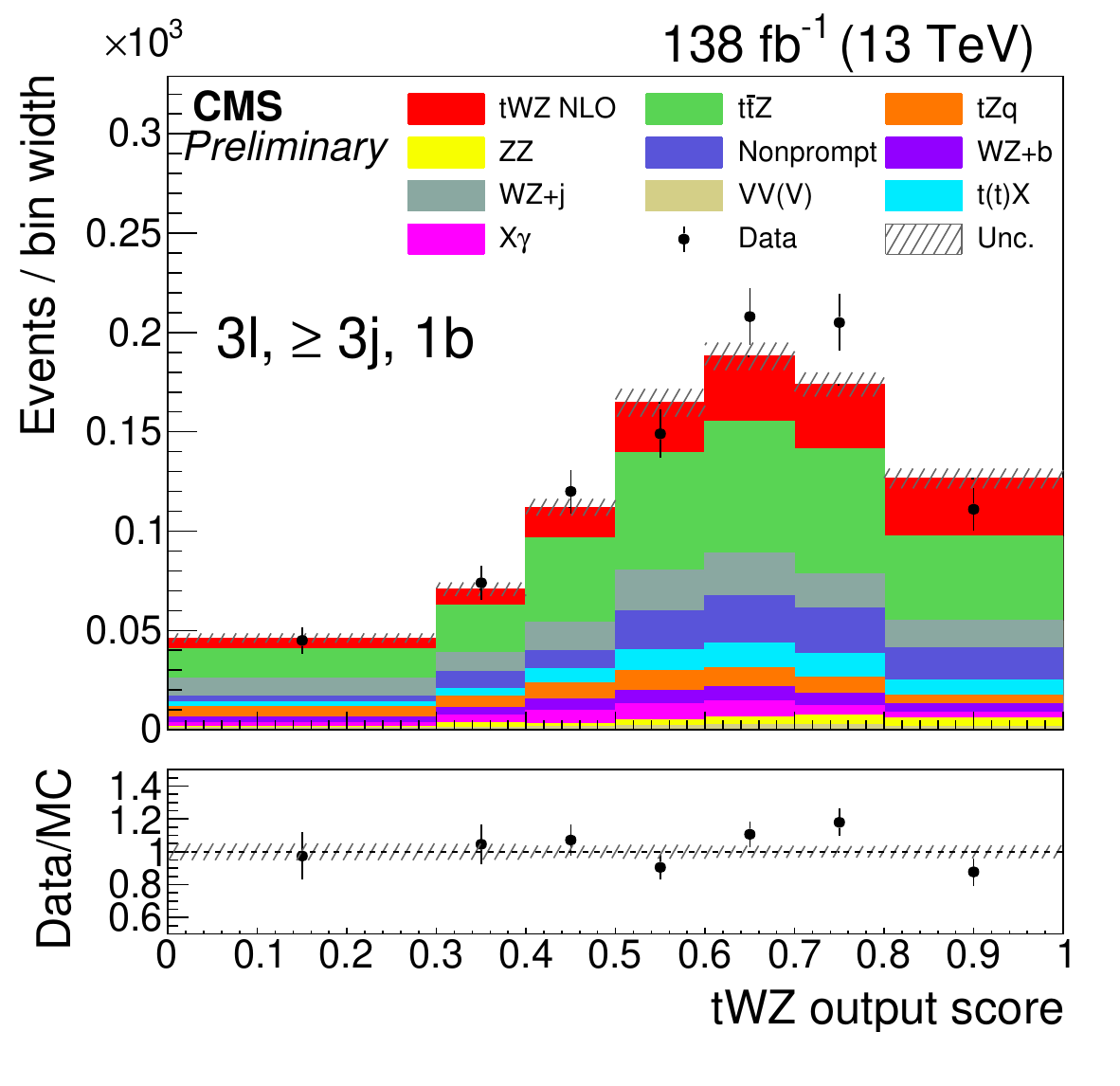}
\includegraphics[width=0.325\textwidth]{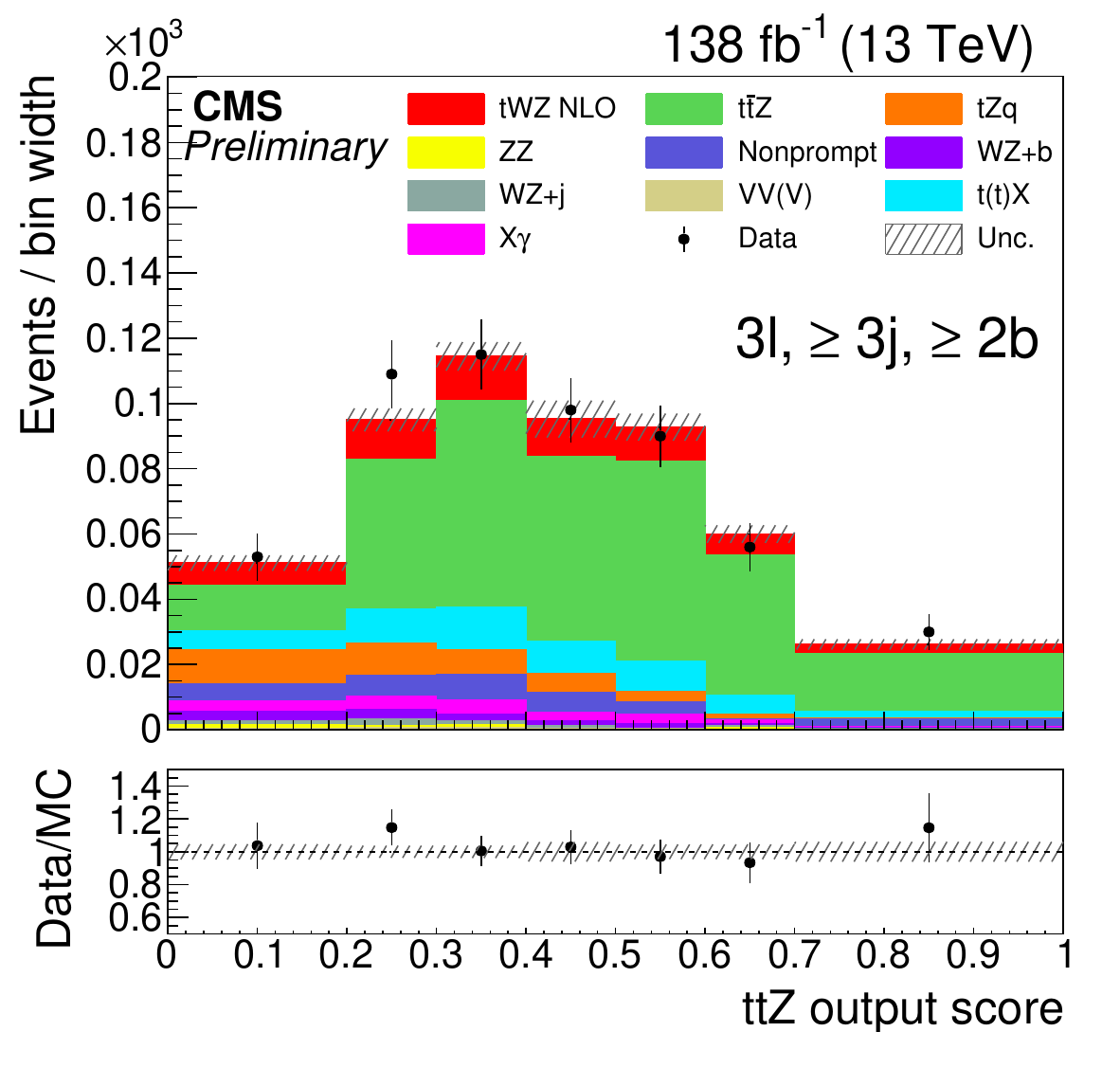}
\caption{Postfit distributions of NN output scores for different SRs in the search for tWZ production from the CMS Collaboration~\cite{CMS:2023dny}.}
\label{fig:twz}
\end{figure}

Contrary to the previous results, the search for the associated production of a top quark, a W boson and a Z boson (tWZ) by the CMS Collaboration with $138 \, \mathrm{fb}^{-1}$ of data~\cite{CMS:2023dny} targets the W-associated single top production mode. Multiple decay modes of the combined tWZ system are considered, resulting in multiple SRs and CRs with different numbers of leptons and (b-tagged) jets. Binary and multiclassification NNs are employed to target tWZ production, as well as the dominant background process, the associated production of a top quark pair with a Z boson (ttZ). With a simultaneous fit in all SRs and CRs a signal is extracted with a significance of $3.5\sigma$ ($1.4\sigma$ expected) and the cross section is measured as
\beq
\sigma_{\mathrm{tWZ}} = 0.37 \pm 0.05 \,\mathrm{(stat)} \pm 0.10 \,\mathrm{(syst)} \,\mathrm{pb} \,,
\eeq{eq:tWZ_cms}
which is compatible with the SM prediction within $2.1\sigma$. This is the first evidence for tWZ production at the LHC. The normalization of the ttZ background is the most dominating systematic uncertainty in this search. Postfit distributions in three of the SRs are shown in Fig.~\ref{fig:twz}.

\section{Summary}
Despite the low cross section of single top quark production in association with vector bosons these processes have been successfully measured by the experiments at the LHC. While it is already possible to measure tZq production differentially, the statistical uncertainties in all the measurements are still large. Therefore, it will be important to measure these processes also at Run 3 of the LHC, especially with the observed deviations with respect to the SM predictions.





\bibliography{eprint}{}
\bibliographystyle{ieeetr}

\end{document}